\documentclass[onecolumn,showpacs,preprintnumbers,superscriptaddress,amsmath,amssymb]{revtex4}
\usepackage[dvips]{graphicx}
\usepackage{latexsym,amssymb,amsmath,epsfig,bm,times,psfrag}
\usepackage{color}
\usepackage[bookmarksnumbered,bookmarksopen,colorlinks,citecolor=red,linkcolor=blue]{hyperref}
\usepackage{epstopdf}
\usepackage[capitalize]{cleveref}
\newcommand{\eq}[1]{Eq.~(\ref{#1})}
\newcommand{\sect}[1]{Sec.~\ref{#1}}

\renewcommand{\part}{{\rm part}}

\renewcommand{\vec}{\boldsymbol}
\newcommand{\be}{\begin{equation}}
\newcommand{\ee}{\end{equation}}
\newcommand{\bear}{\begin{eqnarray}}
\newcommand{\eear}{\end{eqnarray}}
\newcommand{\ba}{\begin{array}}
\newcommand{\ea}{\end{array}}

\begin{document}

\title{Dynamical quark mass and finite volume effects in the Dyson-Schwinger Equations}

\author{Li-Juan Zhou}
\email{zhoulj@gxust.edu.cn}
\affiliation{School of Science, Guangxi University of Science and Technology, Liuzhou, 545006, China}

\author{De-Xian Wei}
\affiliation{School of Science, Guangxi University of Science and Technology, Liuzhou, 545006, China}

\author{Zhong-Yi Liu}
\affiliation{School of Science, Baise University, Baise, 533000, China}

\author{Hong-Wei Zhong}
\affiliation{School of Science, Guangxi University of Science and Technology, Liuzhou, 545006, China}

\date{\today}

\begin{abstract}
Within the framework of Dyson-Schwinger equations(DSEs) and by means of the Multiple Reflection Expansion approximation, we study the finite volume effects of the constituent quark mass in a strong external magnetic field. Since the magnetic field has influence on the coupling constant, the coupling constant controls the strength of strongly interaction in QCD, so we adopt the magnetic-field-dependent running coupling constant in simulation. The results show that in addition to the magnetic field, the masses of constituent quarks also have a significant dependence on the volume and the running coupling constant. The model behaves close to the infinite volume limit for large size, but the effect of the finite volume is significant when the system size $R$ is about $2-6$ fm.
The finite volume effects and the magnetic-field-dependent running coupling constant have considerable influence on the phase transition.
\end{abstract}


\maketitle


\section{Introduction}
\label{sec:sec1}

Studying the QCD phase transition is always challenging in high-energy physics since the research may reveal the nature of the early universe matter evolution~\cite{Braun:2009RMP,Phili:2013PNP,Gupta:2011SCI}. The transitions include chiral phase transition and the confinement
transition. Many physical parameters affect the QCD phase transition, where besides the temperature and chemical potential, magnetic field and system volume also affects the transition. Since the system generated in heavy-ion collision experiments has finite volume which depends on the size of the colliding nuclei, the collision center of mass energy and the centrality of collision ~\cite{Steph:1999PRD,Steph:2011JPG,Bha:2013PRD}. In Ref.~\cite{Bha:2007Nat,Pal:2011JPG,NPA:2005} estimated the quark-gluon plasma (QGP) system produced at the relativistic heavy ion collision (RHIC) could have sizes between 2 fm and 10 fm. So a comprehension of finite volume effects is very important for the relativistic heavy ion collision.

Theoretical physicists have put forward many theoretical models to study the finite volume effects on strongly interacting systems, including the Nambu-Jona-Lasinio (NJL) model ~\cite{Bha:2015PRC,Koh:2016NPB}, the quark-meson model~\cite{Bra:2005PRD,Bra:2006PRD}, the non-interaction bag model~\cite{Elze:1986PLB} and Dyson-Schwinger Equations (DSEs)~\cite{Sama:2018JPG,Abreu:2017PRC,Steph:2002PRD,Farias:2019tei,Farias:2020tei,Farias:2021tei,Farias:2023tei}. Most of these studies, for the sake of convenience, the systems are considered as a cub, antiperiodic boundary condition(APBC) as well as periodic boundary condition(PBC) are used. However, if the volume of fireball is small, we must take the system size and shape into account. This lead us to consider a more realistic approach to study the QCD phase transition. In this work, we adopt the Multiple Reflection Expansion (MRE) approximation ~\cite{Balian:1970MRE,Madsen:1994MRE,Kiri:2003MRE,Kiri:2005MRE}, which consider the surface contribution and curvature contribution.

With the development of experiments, very strong magnetic fields have been generated in noncentral heavy-ion collision experiments. Since quarks are electrically charged and can be coupled to magnetic field, so the magnetic field has great influence on the QCD phase transition. A natural question that emerges is what is the effect of the magnetic field produced in collisions on QCD phase transition in a finite volume system. This question has been partially studied in Ref~\cite{Nisha:2023vam,Prabal:2023QCD,Somenath:2023eof}. In addition, it has been confirmed that the magnetic field also exerts influence on the coupling constant~\cite{Miransky:2002prd}. People have worked hard to construct a magnetic-field-dependent running coupling constant~\cite{Farias:2014prc,Ferrer:2015qaa,Farias:2017tei,Tavares:2021nsm}. We believe that, in the presence of a magnetic field, constructing a more realistic framework that is able to meet with the physical coupling strength of QCD is reasonable and necessary. Nevertheless, the incorporation of QCD properties and the substantial computational challenges associated with solving the coupled equations remain significant hurdles.

In our previous works, we have studied the condensate and magnetization phenomena in strong external magnetic field by use of the DSEs, the results showed that the such condensate and magnetization phenomena are dependent on the magnitude of magnetic field~\cite{wei2023quark, Zhou:2014tdo}. As we known, the system created in RHIC exists in finite size, rather than thermodynamical limit. To determine the freeze-out parameters in experiment one has to take the finite size and the running coupling constant into account. Therefore, in this paper, we adopt a different approach based on the DSEs, utilizing MRE formalism to investigate the finite volume effects of strongly interacting matter with a magnetic-field-dependent running coupling constant. We studies its influences on quark condensate and the constituent quark mass. Compared to other methods, it is more natural to study the QGP fireball by means of MRE, which describes a sphere rather than a cube.

The paper is organized as follows: In \sect{sec:sec2} we give a briefly introduction to the DSEs, the MRE formalism and the magnetic-field-dependent running coupling constant. By solving the DSEs in finite volume with the magnetic-field-dependent running coupling constant, we show the numerical results in \sect{sec:sec3}.
Finally, we summarize the main results in \sect{sec:sum}.

\section{Theory model}
\label{sec:sec2}

\subsection{Dyson-Schwinger Equations}

In this section, we will briefly review the formula of DSEs. We start from the quark propagator.
The DSEs in position space and with local interaction is given by
\begin{eqnarray}\label{dse:def101}
S^{-1}(x,y) &=& S_{0}^{-1}(x,y)+\Sigma(x,y),
\end{eqnarray}
where $S^{-1}$ is the inverse of dressed quark propagator,
$S_{0}^{-1}$ is the inverse of free quark propagator, and the quark self energy reads
\begin{eqnarray}\label{dse:def102}
\Sigma(x,y) &=& ig^{2}C_{F}\gamma^{\mu}S(x,y)\Gamma^{\nu}(y)D_{\mu\nu}(x,y),
\end{eqnarray}
with $g$ is the coupling constant of strong interaction, $C_{F} = (N_{c}^{2}-1)/N_{c}$, $\Gamma^{\nu}$ is the dressed quark-gluon vertex, and $D_{\mu\nu}$ denotes the gluon propagator in Landau gauge.
We expand \eq{dse:def101} in terms of Ritus transformation functions, which is a substitute for the usual Fourier exponential factor $e^{ip\cdot x}$.

After some calculating, we can obtain the dressing functions, which are given by~\cite{Mueller:2014dqm}
\begin{eqnarray}\label{dse:def203}
A_{0}(p)|_{L_{p}=L} &=& Z_{2}m_{f} +C_{1}\int_{q}\left\{\left(\frac{A_{0}(q)}{A_{0}^{2}(q)+A_{\parallel}^{2}(q)q_{\parallel}^{2}+ A_{\perp}^{2}(q)q_{\perp}^{2}}\right)\Bigg|_{L_{q}=L} \cdot e^{-\frac{k_{\perp}^{2}}{|2eB|}}G_{1}(k^{2})D(k^{2})\Gamma(k^{2})\right\} ~\nonumber\\
 &+& \frac{C_{2}}{p_{\parallel}^{2}}\frac{2}{\tau(L)}\sum_{L_{q}=L\pm1}\int_{q}\left\{\left(\frac{A_{0}(q)}{A_{0}^{2}(q)+A_{\parallel}^{2}(q)q_{\parallel}^{2}+ A_{\perp}^{2}(q)q_{\perp}^{2}}\right)\Bigg|_{L_{q}}
 \cdot e^{-\frac{k_{\perp}^{2}}{|2eB|}}G_{2}(k^{2})D(k^{2})\Gamma(k^{2})\right\},
\end{eqnarray}

\begin{eqnarray}\label{dse:def204}
A_{\parallel}(p)|_{L_{p}=L} &=& Z_{2} -\frac{C_{1}}{p_{\parallel}^{2}}\int_{q}
 \left\{\left(\frac{A_{\parallel}(q)}{A_{0}^{2}(q)+A_{\parallel}^{2}(q)q_{\parallel}^{2}+ A_{\perp}^{2}(q)q_{\perp}^{2}}\right)\Bigg|_{L_{q}=L}\cdot e^{-\frac{k_{\perp}^{2}}{|2eB|}}G_{3}(p,q,k^{2})D(k^{2})\Gamma(k^{2})\right\} ~\nonumber\\
 &+& \frac{C_{2}}{p_{\parallel}^{2}}\frac{2}{\tau(L)}\sum_{L_{q}=L\pm1}\int_{q}\left\{\left(\frac{A_{\parallel}(q)}{A_{0}^{2}(q)+A_{\parallel}^{2}(q)q_{\parallel}^{2}+ A_{\perp}^{2}(q)q_{\perp}^{2}}\right)\Bigg|_{L_{q}}\cdot e^{-\frac{k_{\perp}^{2}}{|2eB|}}G_{4}(p,q,k^{2})D(k^{2})\Gamma(k^{2})\right\},
\end{eqnarray}

\begin{eqnarray}\label{dse:def205}
A_{\perp}(p)|_{L_{p}=L} &=& Z_{2} +\frac{C_{1}}{p_{\parallel}^{2}}\int_{q}
 \left\{\left(\frac{A_{\perp}(q)}{A_{0}^{2}(q)+A_{\parallel}^{2}(q)q_{\parallel}^{2}+ A_{\perp}^{2}(q)q_{\perp}^{2}}\right)\Bigg|_{L_{q}=L}\cdot e^{-\frac{k_{\perp}^{2}}{|2eB|}}G_{5}(p,q,k^{2})D(k^{2})\Gamma(k^{2})\right\} ~\nonumber\\
 &-& \frac{C_{2}}{p_{\parallel}^{2}}\frac{2}{\tau(L)}\sum_{L_{q}=L\pm1}\int_{q}\left\{\left(\frac{A_{\perp}(q)}{A_{0}^{2}(q)+A_{\parallel}^{2}(q)q_{\parallel}^{2}+ A_{\perp}^{2}(q)q_{\perp}^{2}}\right)\Bigg|_{L_{q}}\cdot e^{-\frac{k_{\perp}^{2}}{|2eB|}}G_{6}(p,q,k^{2})D(k^{2})\Gamma(k^{2})\right\},
\end{eqnarray}
where
\begin{eqnarray}\label{dse:def206}
   \begin{cases}
        \tau(L)=2, & L=0\\
        \tau(L)=4, & L>0   .
    \end{cases}
\end{eqnarray}
We make the following identifications in \eq{dse:def203}-\eq{dse:def205}
\begin{eqnarray}\label{dse:def211}
\int_{q} &=& \int\frac{d^{2}q_{\parallel}}{(2\pi)^{4}}\int_{-\infty}^{\infty}dq_{2}dk_{1}, ~\nonumber\\
C_{1} &=& Z_{1f}g^{2}C_{F}, ~~~~ C_{2} = g^{2}C_{F}, ~\nonumber\\
G_{1}(k^{2}) &=& 2-\frac{k_{\parallel}^{2}}{k^{2}}, ~~~~ G_{2}(k^{2})= 2-\frac{k_{\perp}^{2}}{k^{2}}, ~\nonumber\\
G_{3}(p,q,k^{2}) &=& p_{\parallel}q_{\parallel}cos(\varphi)\frac{k_{\parallel}^{2}}{k^{2}} -\frac{2[p_{\parallel}q_{\parallel}cos(\varphi)-p_{\parallel}^{2}][q_{\parallel}^{2}-p_{\parallel}q_{\parallel}cos(\varphi)]}{k^{2}}, ~\nonumber\\
G_{4}(p,q,k^{2}) &=& \left(2-\frac{k_{\perp}^{2}}{k^{2}}\right)p_{\parallel}q_{\parallel}cos(\varphi), ~\nonumber\\
G_{5}(p,q,k^{2}) &=& \left(2-\frac{k_{\parallel}^{2}}{k^{2}}\right)p_{\perp}q_{\perp}, ~\nonumber\\
G_{6}(p,q,k^{2}) &=& \left(\frac{k_{1}^{2}-k_{2}^{2}}{k^{2}}\right)p_{\perp}q_{\perp}, ~\nonumber\\
cos(\varphi) &=& \frac{\vec{p}_{\parallel}\vec{q}_{\parallel}}{|p_{\parallel}||q_{\parallel}|}.  ~\nonumber
\end{eqnarray}

As we know, solving DSEs in a strong magnetic field is quite difficult. For simplicity,
we choose the following forms for the quenched gluon propagator $D(k^{2})$ and the vertex dressing function $\Gamma(k^{2})$~\cite{Mueller:2014dqm}:
\begin{eqnarray}\label{dse:def212}
D(k^{2}) &=& \frac{1}{k^{2}}\frac{k^{2}\Lambda^{2}}{(k^{2}+\Lambda^{2})^{2}} \left\{\left(\frac{c}{k^{2}+a\Lambda^{2}}\right)^{b} +\frac{k^{2}}{\Lambda^{2}}\left[\frac{\beta\alpha(\mu)log[\frac{k^{2}}{\Lambda^{2}}+1]}{4\pi}\right]^{\gamma}\right\},
\end{eqnarray}

\begin{eqnarray}\label{dse:def212-1}
\Gamma(k^{2}) &=& \frac{d_{1}}{d_{2}+k^{2}}+\frac{k^{2}}{k^{2}+\Lambda^{2}}\left[\frac{\beta\alpha(\mu)log[\frac{k^{2}}{\Lambda^{2}}+1]}{4\pi}\right]^{2\delta}. 
\end{eqnarray}
here parameters are $a$=0.60,~$b$=1.36,~$\Lambda$=1.4~GeV,~$c$=11.5~GeV$^{2}$,~$\beta$=22/3,~$\gamma$=-13/22, ~$d_{1}$=7.9~GeV$^{2}$,~$d_{2}$=0.5~GeV$^{2}$,~$\delta$=-18/88.
The quark mass renormalization factor $Z_{2}$ is determined in the renormalisation process, the renormalization factor of the quark-gluon vertex is denoted by $Z_{1f}$, and satisfies an approximate Slavnov-Taylor identity in the infrared and the correct ultraviolet running from resummed perturbation theory.

Substituting \eq{dse:def212}-\eq{dse:def212-1} into \eq{dse:def203}-\eq{dse:def205} one can solve the equations numerically, and get the quark dressing function
$A_{0}$, $A_{\parallel}$ and $A_{\perp}$. Noting that only $A_{0}$ and $A_{\parallel}$ contribute to the lowest Landau level.

\subsection{Finite volume effects: inclusion of MRE}

\eq{dse:def203}-\eq{dse:def205} describes systems in infinite volume. It is meaningful to generalize the previous results to the systems of finite volume, since quark matter created in RHIC exists in finite size, rather than thermodynamical limit. In the present work we consider the finite volume effects in DSEs by means of MRE formalism.

In the MRE formalism, the modified density of states of a finite spherical droplet is given by ~\cite{Balian:1970MRE,Madsen:1994MRE,Kiri:2003MRE,Kiri:2005MRE}
\begin{eqnarray}\label{dse:MRE1}
\rho_{MRE}(q,m_{f},R)=1+\frac{6\pi^{2}}{qR}f_{S,f}(q,m_{f})+\frac{12\pi^{2}}{(qR)^{2}}f_{C,f}(q,m_{f})\cdot\cdot\cdot,
\end{eqnarray}
here $m_{f}$ is mass of quarks, and $R$ is the radius of sphere. The higher order terms in $\frac{1}{R}$ corresponds to the ellipsis, for simply, which are neglected here.
\begin{eqnarray}\label{dse:MRE2}
f_{S,f}(q,m_{f})=-\frac{1}{8\pi}(1-\frac{2}{\pi}\arctan\frac{q}{m_{f}}),
\end{eqnarray}
and
\begin{eqnarray}\label{dse:MRE3}
f_{C,f}(q,m_{f})=\frac{1}{12\pi^{2}}[1-\frac{3q}{2m_{f}}(\frac{\pi}{2}-\arctan\frac{q}{m_{f}})],
\end{eqnarray}
where$f_{S,f}(q,m_{f})$ and $f_{C,f}(q,m_{f})$ corresponding to the surface contribution and curvature contribution respectively.

When finite volume effects are considered, the integrals of \eq{dse:def203}-\eq{dse:def205} replaced by:
\begin{eqnarray}\label{dse:MRE4}
\int^{\infty}_{0}\frac{d^{3}q}{(2\pi)^{3}}\cdot\cdot\cdot\rightarrow \int^{\infty}_{\Lambda_{IR}}\frac{d^{3}q}{(2\pi)^{3}}\rho_{MRE}\cdot\cdot\cdot.
\end{eqnarray}

When $m_{f}\neq0$, the MRE density of states would become negative leading to a unphysical results. In order to remove the unphysical values, we introduce an infrared cutoff $\Lambda_{IR}$, which is the largest solution of the equation $\rho_{MRE}(q,m_{f},R) = 0 $ with respect to the momentum $q$. Although the finite volume effect was chosen, it should not be corrected for the vertex function at $T$=0 and $\mu$=0.  That is, the volume effect is a thermodynamic effect that has the same basis as temperature, chemical potential ~\cite{Sarkar:2025xpv}.
When $m_{f}=0$, the MRE density of states is given by ~\cite{Madsen:1994MRE}
\begin{eqnarray}\label{dse:MRE1}
\rho_{MRE}(q,m_{f},R)= 1-\frac{1}{2(qR)^{2}}.
\end{eqnarray}

\subsection{The magnetic-field-dependent running coupling constant}

As is known the magnetic field has influence on the running coupling constant, the coupling constant $g$ in \eq{dse:def203}-\eq{dse:def205} controls the strength of strongly interaction in QCD. So the magnetic-field ($eB$)-dependent running coupling constant will have important influence on the phase of QCD. In this work we adopt the $eB$-dependent running coupling constant ~\cite{Farias:2014prc}
\begin{eqnarray}\label{dse:GB}
g_{\mathrm{II}}=\frac{g_\mathrm{I}}{1+D_{1}\ln(1+D_{2}\frac{eB}{\Lambda^{2}_{QCD}})},
\end{eqnarray}
where $\Lambda^{2}_{QCD}$=200~MeV$^2$, $g_\mathrm{I}$ is the coupling constant at $eB=0$, the free parameters $D_{1}$ and $D_{2}$ are fixed to obtain reasonable results of the lattice average$(\Sigma_{u}+\Sigma_{d})/2$ ~\cite{Ding:2022cca}. In our work\cite{wei2023quark}, the subtracted quark condensate is given by
\begin{eqnarray}\label{qc:def102}
\frac{\left(\Sigma_{u}+\Sigma_{d}\right)}{2} &=& 1-\frac{a_{1}}{eB}\frac{m_{L}}{m_{\pi}^{2}f_{\pi}^{2}} \left\{\left[\langle \bar{u}u\rangle_{eB}-\langle \bar{u}u\rangle_{eB=0}\right]+\left[\langle \bar{d}d\rangle_{eB}-\langle \bar{d}d\rangle_{eB=0}\right]\right\},
\end{eqnarray}
where $m_{L}=6$~MeV, $m_{\pi}=138$~MeV and $f_{\pi}=92.4$~MeV. The $a_{1}/eB$ parameter plays the part of the anomalous magnetic moment factor of the external magnetic-dependent chiral condensate,
which is compared with the AMM effect (noted as $\kappa_{u}$, $\kappa_{d}$) in NJL~\cite{Kawaguchi:2022rot}.

The local quark condensates $\langle \bar{u}u\rangle$ and $\langle \bar{d}d\rangle$ are given by
\begin{eqnarray}\label{qc:def101}
-\langle \bar{q}q\rangle &=& Z_{2}N_{c}\frac{eB}{2\pi^{2}}\sum_{L_{q}=0}^{\infty}\frac{\tau(L_{q})}{2}\int_{0}^{\infty}dq_{\parallel}q_{\parallel}
 \left[\frac{A_{0}(q)}{A_{0}^{2}(q)+A_{\parallel}^{2}(q)q_{\parallel}^{2}+ A_{\perp}^{2}(q)q_{\perp}^{2}}\right]\Bigg|_{L_{q}}.
\end{eqnarray}

\section{Numerical Results}
\label{sec:sec3}

In this section we presents the numerical results for dynamically generated quark condensate, quark mass and magnetization phenomena, we will concentrate on the finite volume effects and magnetic fields.
We using the current quark masses, $m_{u}=6$~MeV, $m_{d}=10$~MeV and $m_{s}=199$~MeV, in simulations. Throughout this paper, our results and analyses are considered on zero temperature ($T=0$).

\begin{figure*}[tp]
\begin{center}
\includegraphics[width=0.50\textwidth]{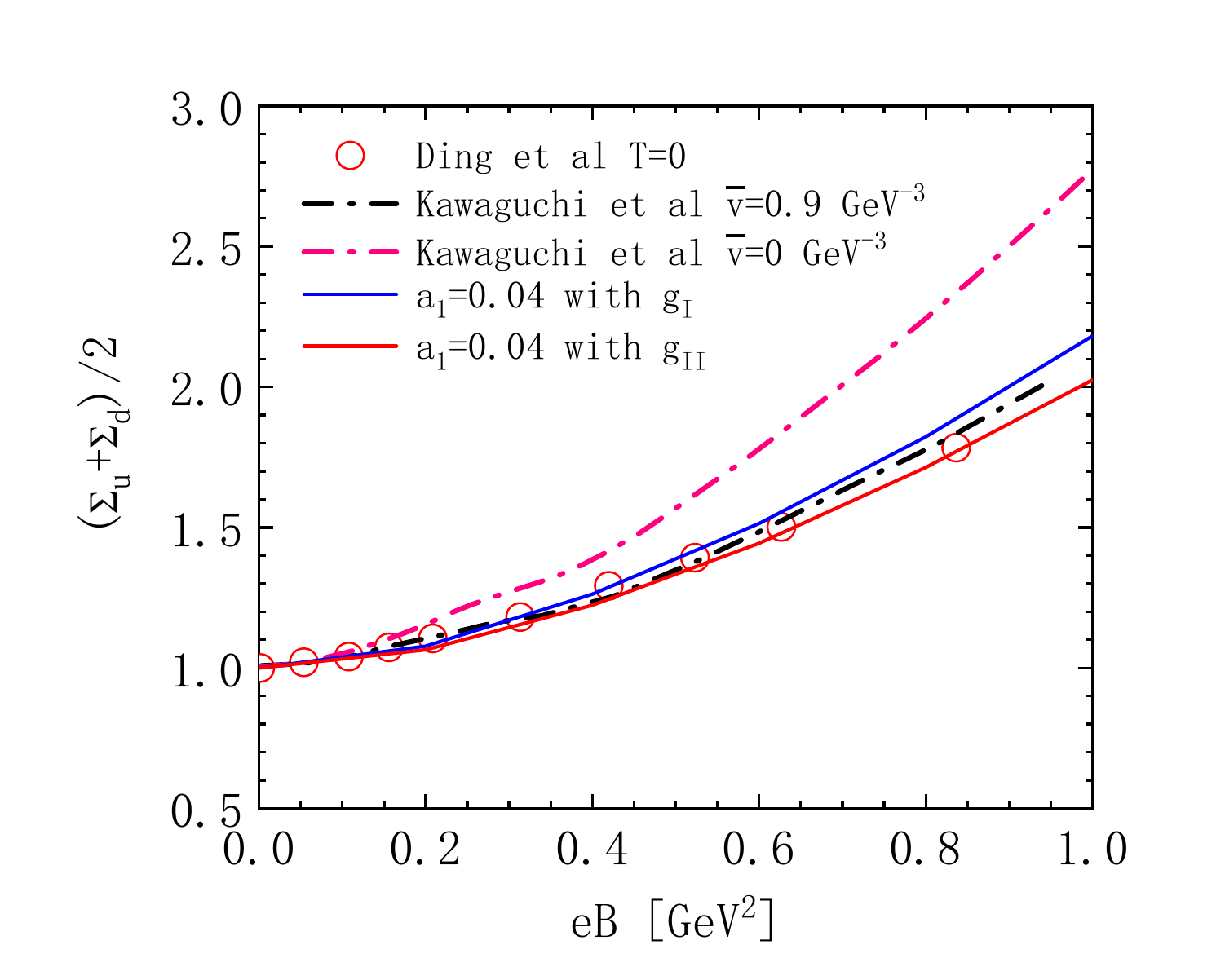}
\caption{(Color online)
The magnetic dependence of the subtracted quark condensate, comparing the results from DSEs, the NJL\cite{Kawaguchi:2022rot} and the lattice QCD data~\cite{Ding:2022cca}.
}
\label{fig1}
\end{center}
\end{figure*}

The Fig.\ref{fig1} shows the magnetic dependence of the subtracted quark condensate $(\Sigma_{u}+\Sigma_{d})/2$. Comparisons with our simulations, the NJL [with AMM ($\bar{v}=0.9$~GeV$^{-3}$) and without AMM ($\bar{v}=0$~GeV$^{-3}$)] results ~\cite{Kawaguchi:2022rot}, and the lattice QCD results~\cite{Ding:2022cca}. We obtain a good fit of the lattice QCD results~\cite{Ding:2022cca} with $D_{1}=2.39$, $D_{2}=0.002515$ and $a_{1}= 0.04$ in \eq{dse:GB}.
In $g_\mathrm{I}$ case, the subtracted quark condensate increases monotonically with the magnetic field using DSEs, that is, the magnetic catalysis (MC) effect.
However, by including the $eB$-dependent running coupling constant (in $g_\mathrm{II}$ case) leads to a suppression of the
magnetic enhancement effect. This phenomenon originates from: (1) The magnetic-field dependence of the effective coupling
manifests as a monotonic decrease with increasing magnetic field ($B$). This diminishing coupling counteracts the naive magnetic
enhancement expected in conventional scenarios. (2) The magnetic field indirectly influences gluon dynamics through its coupling to sea quarks, generating an effect analogous to asymptotic freedom in QCD. This back reaction of gluon constitutes the
primary physical origin of the observed suppression. The interplay between these competing mechanisms provides a consistent
explanation for the suppression.

\begin{figure*}[tp]
\begin{center}
\hspace{-0.3cm}
\includegraphics[width=0.50\textwidth]{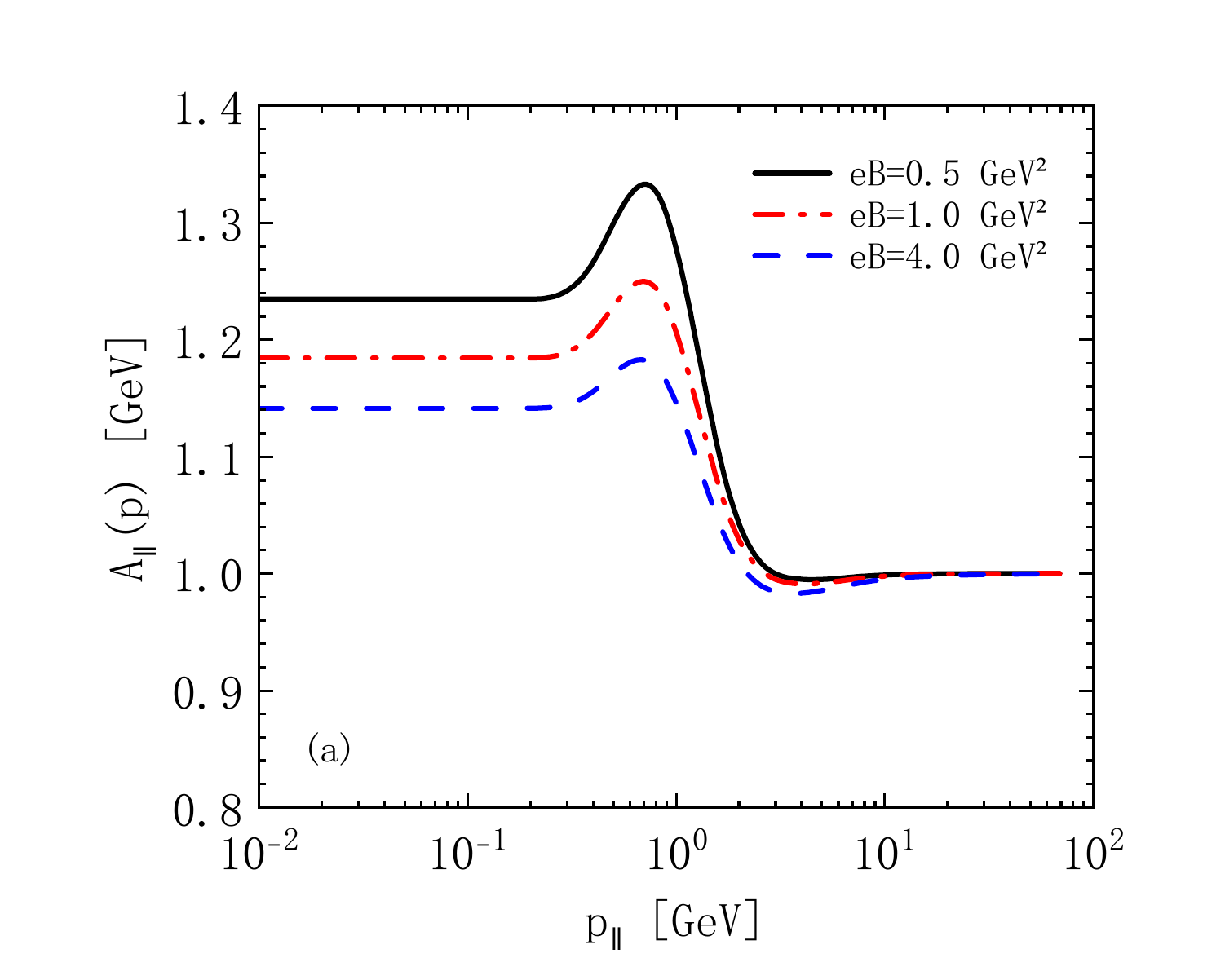}
\hspace{0.01cm}
\includegraphics[width=0.50\textwidth]{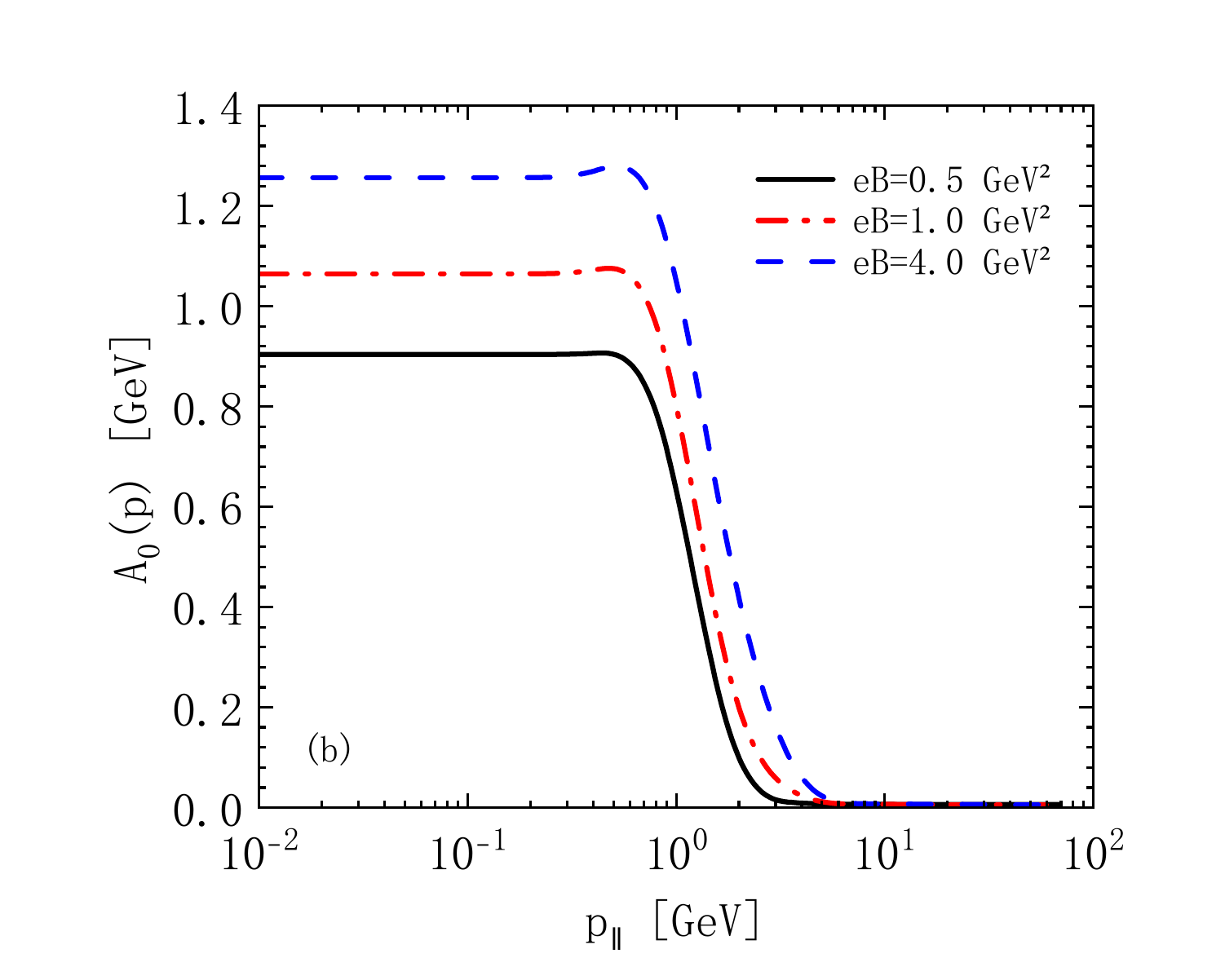}
\caption{(Color online)
The quenched dressing functions $A_{\parallel}$ in (a) and $A_{0}$ in (b) as functions of the momentum $p_{\parallel}$ for $m_{u}$ = 6 MeV at $eB$=0.5, 1.0, 4.0~GeV$^{2}$.
}
\label{fig2}
\end{center}
\end{figure*}

In Fig.\ref{fig2}(a) and (b), we plot the momentum dependence of the quenched dressing functions $A_{\parallel}$ , and $A_{0}$ at different magnetic fields. The calculating results of the quenched dressing functions $A_{\parallel}$ and $A_{0}$ without volume effects as a basic test of current work. In order to demonstrate the finite volume effect, we use the MRE formalism, and we then solve the DSEs in the presence of magnetic fields with the same parameters.

\begin{figure*}[tp]
\begin{center}
\includegraphics[width=0.50\textwidth]{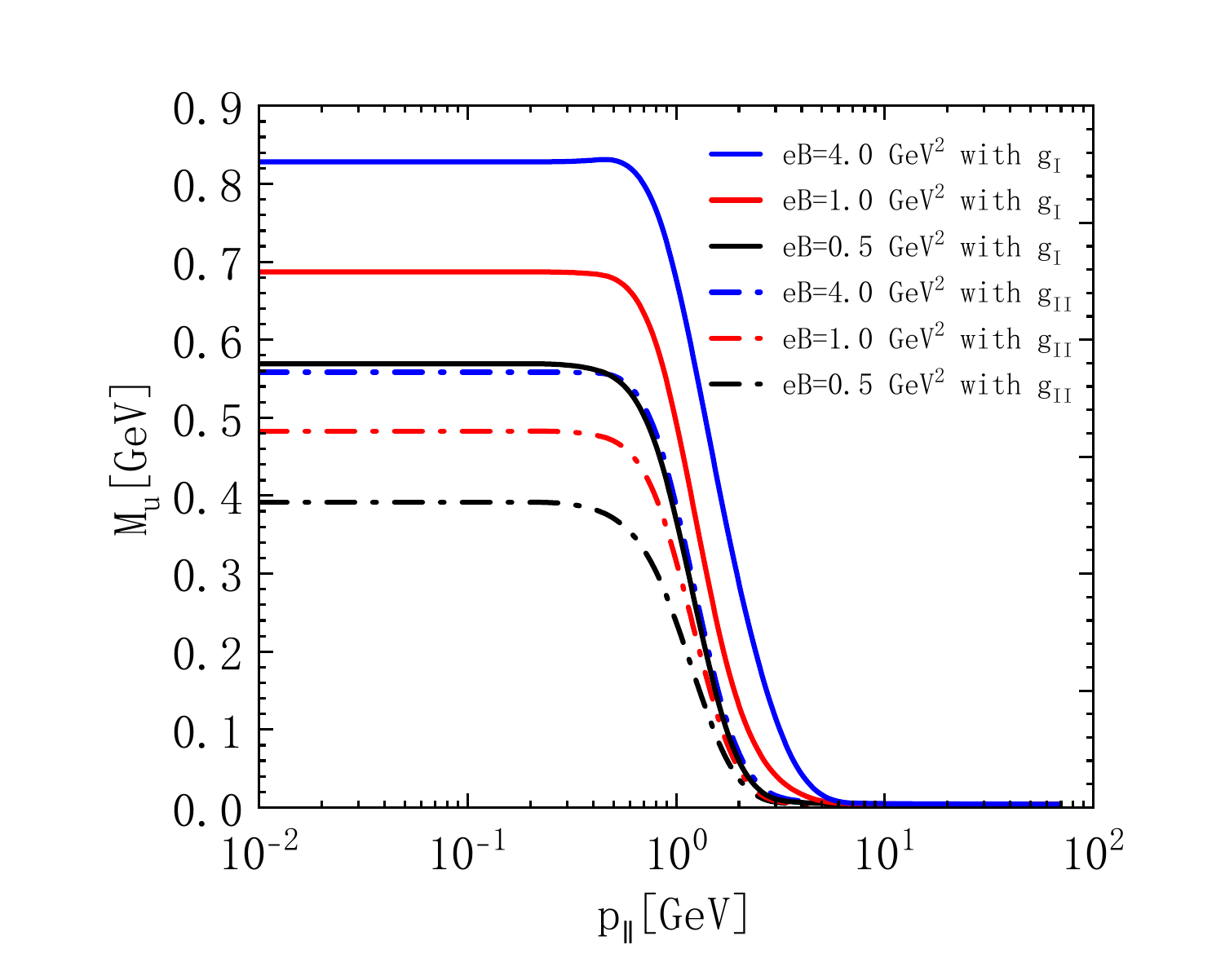}
\caption{(Color online)
Comparision between $g_\mathrm{I}$ and $g_\mathrm{II}$: constituent quark mass $M_{u}$ as a function of the momentum $p_{\parallel}$ for $m_{u}$ = 6 MeV at $eB$=0.5, 1.0, 4.0~GeV$^{2}$.
}
\label{fig3}
\end{center}
\end{figure*}

\begin{figure*}[tp]
\begin{center}
\hspace{-0.3cm}
\includegraphics[width=0.50\textwidth]{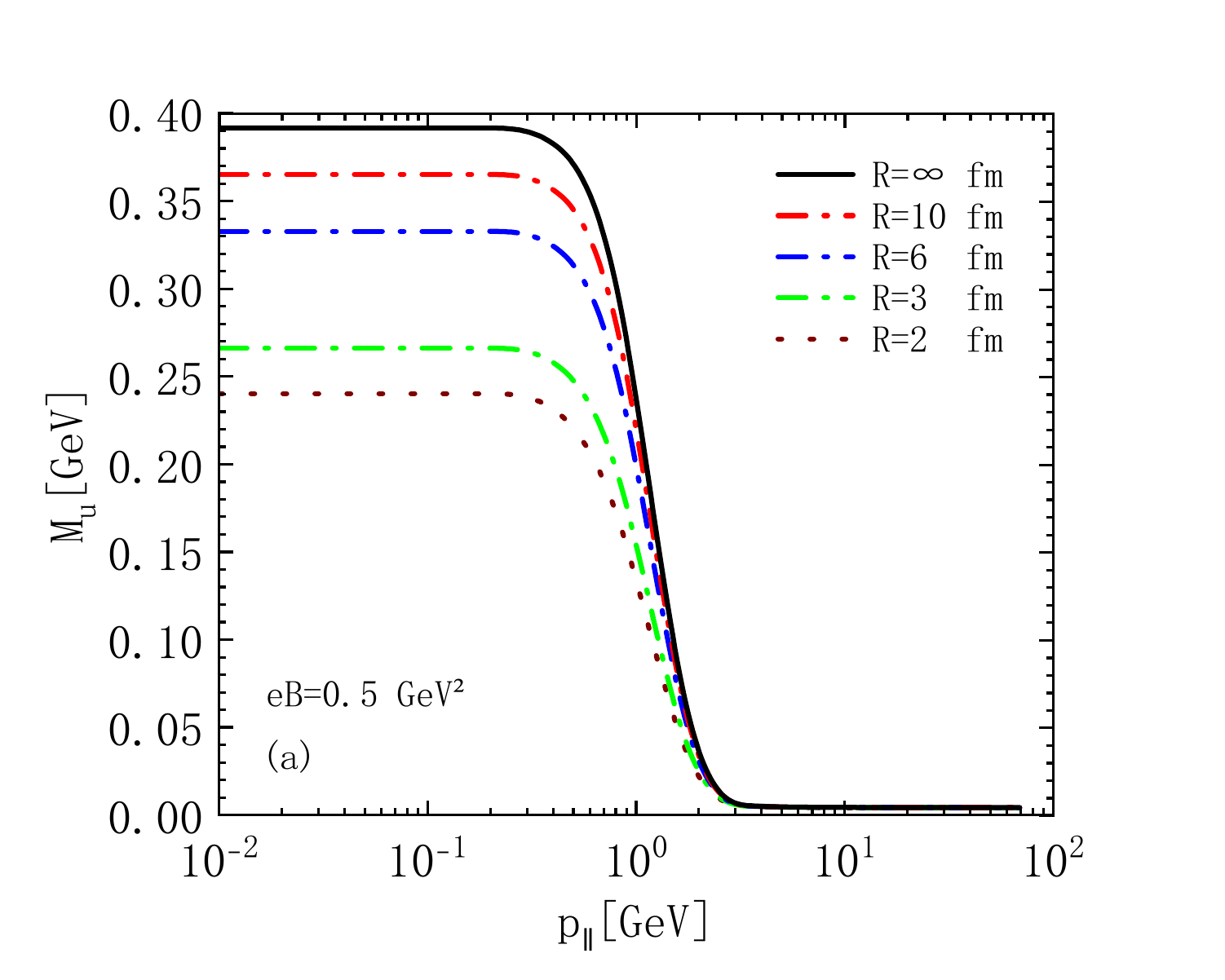}
\hspace{0.01cm}
\includegraphics[width=0.50\textwidth]{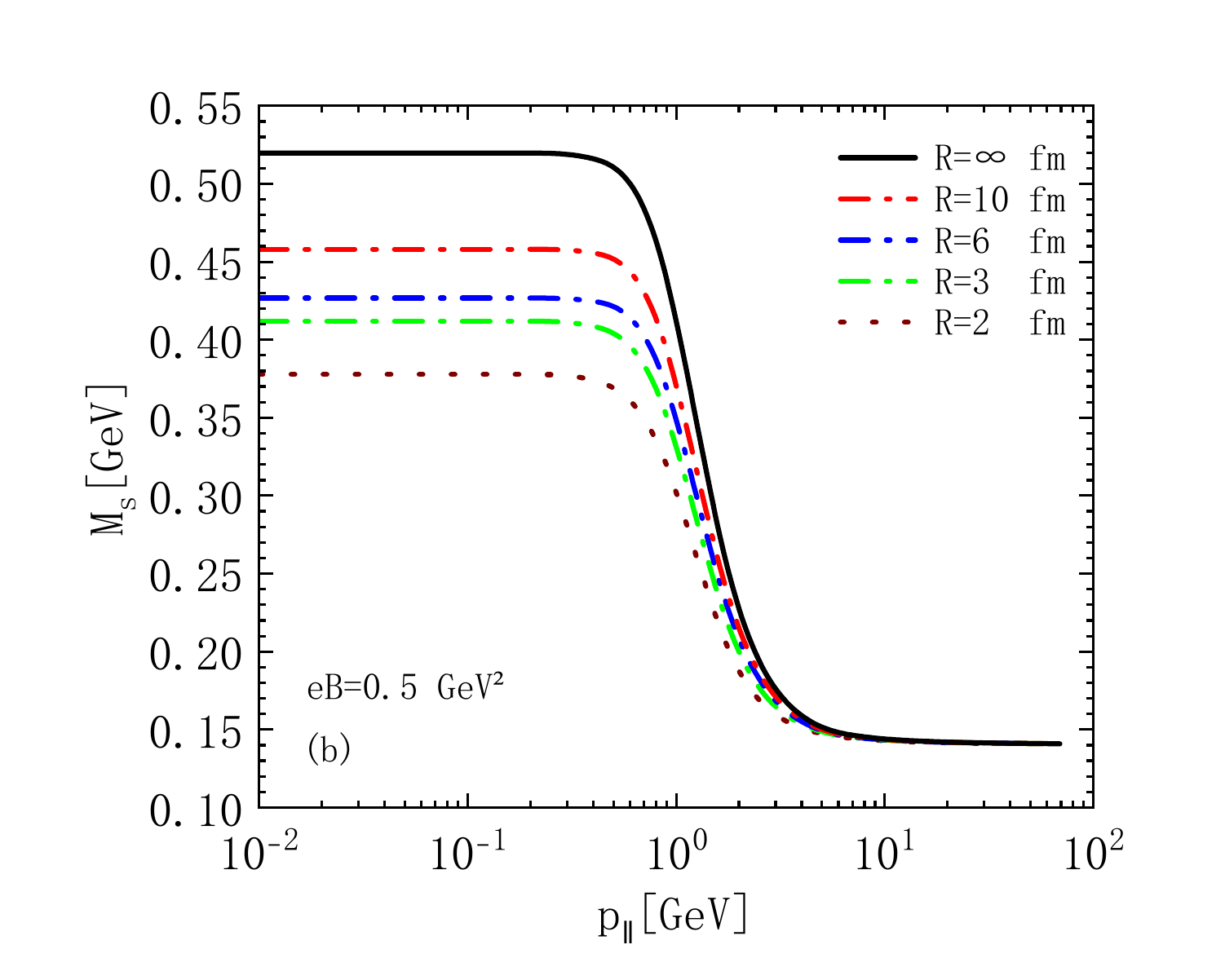}
\caption{(Color online)
Constituent quark mass $M_{u}$ as functions of the momentum $p_{\parallel}$ for $m_{u}$ = 6 MeV in (a) and for $m_{s}$ = 199 MeV in (b) for different radius $R$.
}
\label{fig4}
\end{center}
\end{figure*}

\begin{figure*}[tp]
\begin{center}
\includegraphics[width=0.50\textwidth]{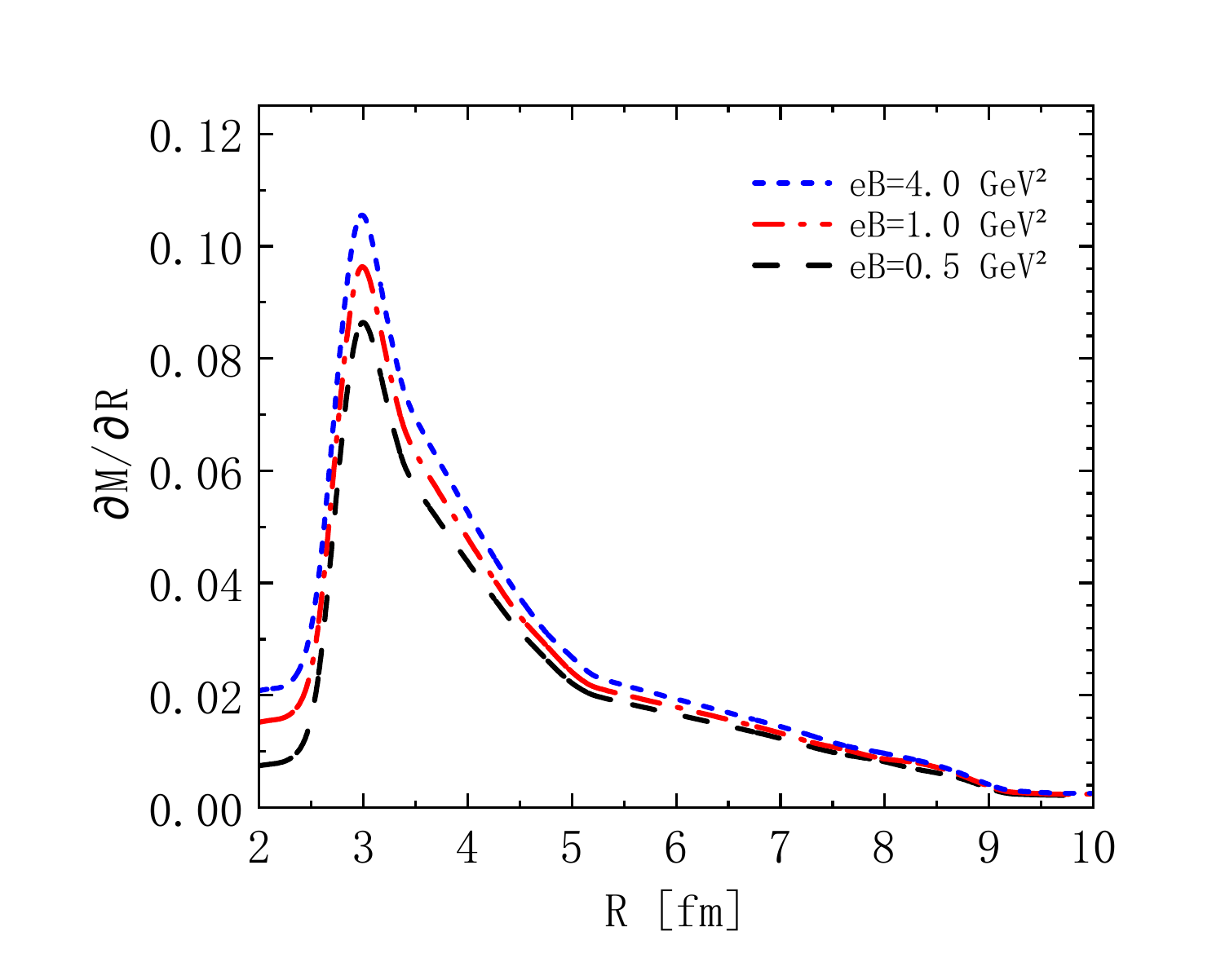}
\caption{(Color online)
The partial derivative respecting to the radius $\partial M/\partial R$ for $m_{u}$ = 6 MeV at $eB$=0.5, 1.0, 4.0~GeV$^{2}$.
}
\label{fig5}
\end{center}
\end{figure*}

\begin{figure*}[tp]
\begin{center}
\includegraphics[width=0.50\textwidth]{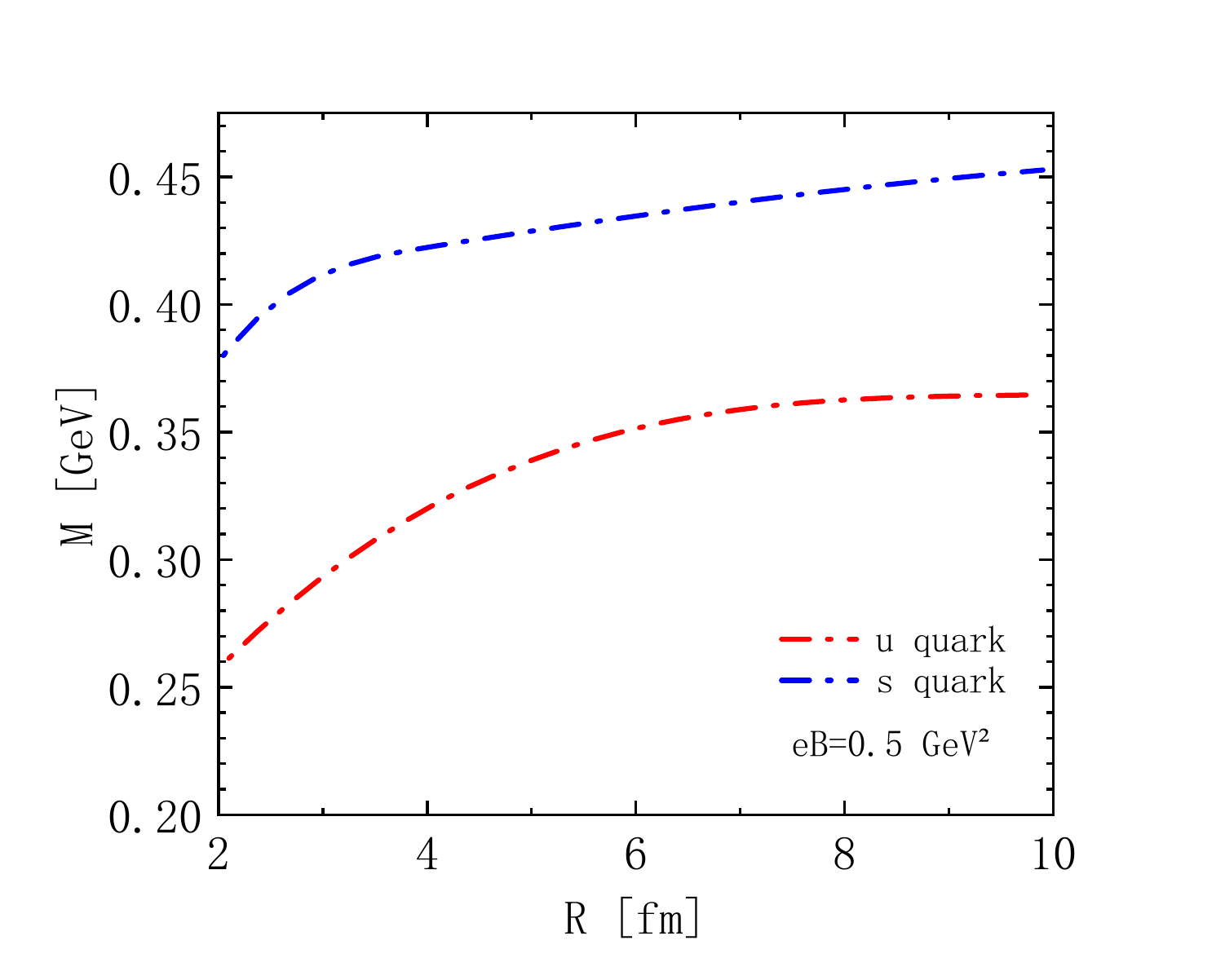}
\caption{(Color online)
Constituent quark mass $M_{u}$ and $M_{s}$ as functions of radius $R$.
}
\label{fig6}
\end{center}
\end{figure*}

Using the solutions of the quark's DSEs, we obtain the constituent quark mass $M=A_{0}/\sqrt{A_{\parallel}^{2}+A_{\perp}^{2} }$. In Fig.\ref{fig3}, the constituent quark mass $M$ is presented as a function of the momentum $p_{\parallel}^{2}$ at different magnetic fields, $eB$=0.5, 1.0, 4.0~GeV$^{2}$, for $g_\mathrm{I}$ and $g_\mathrm{II}$ cases, respectively. We find the constituent quark mass $M$ has a significant dependence on the running coupling constant. As the magnetic field becomes stronger, the constituent quark mass is obviously increase indicative for magnetic catalysis, as showed in $g_\mathrm{I}$ case. Consider the $g_\mathrm{II}$ case, in the small momentum region, the mass of constituent quark $M_{u}$ drops from 687 to 483 MeV and decrease $30\%$ compare with the $g_\mathrm{I}$ case at $eB=1.0$~GeV$^{2}$. As the magnetic field becomes stronger, the constituent quark mass drops even more. In the following study we will concentrate on the finite volume effects in $g_\mathrm{II}$ case.

Fig.\ref{fig4}(a) and (b) is a supplement to Fig.\ref{fig3}, we display the momentum and finite volume dependence of the constituent quark mass for $m_{u}$ = 6 MeV and for $m_{s}$ = 199 MeV based on $g_\mathrm{II}$ at the magnetic field $eB=0.5$ GeV$^{2}$. We find the masses of constituent quark u show strong volume dependence besides magnetic dependence. In the small momentum region, $M$ decreases obviously with the decreases of radius $R$, where the the finite volume effects perform prominently. When the radius decreases from infinity to 2 fm, the mass of u quark drops from 392 MeV to 240 MeV and s quark drops from 519 MeV to 377 MeV. At the big momentum region, $M$ decreases slightly with the decreases of radius $R$. When the radius $R$ is quite small, the constituent quark mass is small enough to close to chiral limit. This indicates that the Dynamical Chiral Symmetry Breaking (DCSB) effects reduce with decreasing volumes. The behavior of $M$ approaches that of infinite volume as the radius is large, such as $R\geq$ 10 fm , where the finite volume effect can be ignored safely. Our calculations indicate that DCSB mechanism is realized in idealized systems where the fireball volume is sufficiently large. A similar conclusion has been addressed in prior studies involving DSEs~\cite{Bernhardt:2021iql}. For a finite volume $V$, if the condition  $V\cdot m \cdot \langle \bar{\Psi}\Psi\rangle \gg \pi $ is satisfied, one can still observe the spontaneous formation of quark condensate and the generation of dynamical mass.
While within the framework of DSEs, thermodynamic quantities are expected to diverge when $R >$ 5 fm~\cite{Xu:2020loz}. This divergence arises due to the associated length scale of the system becoming physically invalid under such conditions.

To find out the finite volume effect which affect the constituent quark mass, we calculate the partial derivative of $M$ with respect to radius $R$, $\partial M/\partial R$ for $m_{u}$ = 6 MeV at $eB$=0.5, 1.0, 4.0~GeV$^{2}$, which shows how the range of radius affects $M$. Form Fig.\ref{fig5}, we can see the constituent quark mass $M$ varies with the radius $R$, and $M$ varies more sharply at stronger magnetic field. Once again, when $R >$ 5 fm, the thermodynamic quantities of the system become invalid~\cite{Xu:2020loz}, leading to an almost zero gradient of mass outside the larger volume.
The location of protrusion in Figure \ref{fig5} indicates a sharp variation in the constituent quark mass $M$ for most cases when the radius is confined within a narrow range, approximately between 2 and 6 fm.
The narrow range can be increased by the magnetic field, which means the strong magnetic field can enhance the finite volume effect.

Fig.\ref{fig6} is a supplement to Fig.\ref{fig4}, we display the constituent quark mass $M_{u}$ and $M_{s}$ varies with radius $R$ at $eB$=0.5~GeV$^{2}$, which show that $M$ decreases as the $R$ decreases, and this is the same tendency to change with magnetic in Fig.\ref{fig4}. So, to some extent, reducing the radius $R$ and reducing the magnetic field have a similar effect on the phase transition. Moreover, when radius $R >$ 6 fm the constituent quark mass $M_{u}$ increases slowly, but $M_{s}$ increases obviously even $R >$ 6 fm.

\section{Summary}
\label{sec:sum}

In summary, we study the finite volume effects of the constituent quark mass in strong magnetic field based on the framework of DSEs. Since a sufficiently strong magnetic field will affect the running coupling constant, so the coupling constant in DSEs is replaced by the $eB$-dependent running coupling constant ($g_\mathrm{II}$) in calculation. In addition, within the MRE formalism, the finite volume effects are considered. The results showed that the constituent quark mass are dependent on the size of fireball besides the magnitude of magnetic field.

(I) With the solutions of the quark's DSEs, we obtain the constituent quark mass $M$ by the quenched ladder approximation in $g_\mathrm{II}$ case.
We find that in addition to magnetic dependence, the constituent quark mass also exhibits a strong running coupling constant dependence. As the magnetic field becomes stronger, $M$ is obviously increase indicative for magnetic catalysis in $g_\mathrm{I}$ case. While the magnetic enhancement is suppressed, and shows the inverse magnetic catalysis in $g_\mathrm{II}$ case. In the small momentum region, the masses of constituent quark $M_{u}$ in the $g_\mathrm{II}$ case is suppressed $30\%$ more than that in $g_\mathrm{I}$ case at $eB=1.0$~GeV$^{2}$, i.g., the value of $M_{u}$ is 687 MeV in $g_\mathrm{I}$ case and drops to 483 MeV in $g_\mathrm{II}$ case. As the magnetic field gets stronger, the constituent quark mass drops even more.

(II) Based on the constituent quark mass in the $g_\mathrm{II}$ case, we use the MRE formalism to consider the finite volume effects. We calculate the numerical results for the radius of fireball from 2 to 10 fm. The calculation results show when the momentum is small, $M$ decreases obviously with the decreases of radius $R$, where the finite volume effects perform prominently. When the momentum is big, $M$ decreases slightly with the decreases of radius $R$. The behavior of $M$ approaches that of infinite volume as the radius is large, such as $R\geq$ 10 fm, where the finite volume effect can be ignored safely. To find out the finite volume effect which affect the constituent quark mass, we calculate the partial derivative of $M$ with respect to radius $R$. we can see the constituent quark mass varies more sharply within a narrow range, approximately between 2 fm and 6 fm.

(III) What we shall pay attention to here is that the MRE formalism is valid only when the system is not sufficiently small size, since the MRE density of states would become negative for too small $R$. In summary, our calculation already shows that the finite volume effects and the magnetic-field-dependent running coupling constant have considerable influence on the phase transition.


\section*{Acknowledgements}
L.-J. Zhou has been supported by the National Natural Science Foundation of China Grant No.~11865005, the Natural Science Foundation of Guangxi (China) Grant No.~2025GXNSFAA069552.
D.-X. Wei has been supported by the National Natural Science Foundation of China Grant No.~12105057,
the Natural Science Foundation of Guangxi (China) Grant No.~2023GXNSFAA026020.

\end{document}